\documentclass[pre,aps,amsfonts,amssymb,multicol,epsfig,graphics, referee]{revtex4}

\def \beq{\begin{equation}}         \def \eeq{\end{equation}}
\def \beqa{\begin{eqnarray}}        \def \eeqa{\end{eqnarray}}
\def \bea{\begin{array}}        \def \eea{\end{array}}
\def\nomb{\nonumber}
\def \abs#1{\left| #1 \right|}

\def\bio#1#2#3{{Biophys. J. }{\bf #1}, #2 (#3)}

\def\jpc#1#2#3{{J. Phys. Chem. }{\bf #1}, #2 (#3)}

\def\mol#1#2#3{{Macromolecules }{\bf #1}, #2 (#3)}

\def\nats#1#2#3{{Nature struct. Biol. }{\bf #1}, #2 (#3)}
\def\pnas#1#2#3{{Proc. Natl. Acad. Sci. USA }{\bf #1}, #2 (#3)}
\def\pre#1#2#3{{Phys. Rev. E }{\bf #1}, #2 (#3)}

\def\prl#1#2#3{{Phys. Rev. Lett. }{\bf #1}, #2 (#3)}
\def\sci#1#2#3{{Science }{\bf #1}, #2 (#3)}

\usepackage{amsmath}
\usepackage{epsfig}
\begin{document}

\title{Maximum-entropy calculation of end-to-end distance distribution of 
force stretching chains}
\author{Luru Dai}
\email[]{dailuru@itp.ac.cn}
\affiliation{Institute of Theoretical Physics, The Chinese
Academy of Sciences, P. O. Box 2735, Beijing 100080, China}
\author{Fei Liu}
\affiliation{Institute of Theoretical Physics, The Chinese 
Academy of Sciences, P. O. Box 2735, Beijing 100080, China}
\author{Zhong-can Ou-Yang}
\affiliation{Institute of Theoretical Physics, The Chinese 
Academy of Sciences, P. O. Box 2735, Beijing 100080, China}
\affiliation{Center for Advanced Study, Tsinghua University, Beijing 100084, 
China}


\date{\today}

\begin{abstract}
Using the maximum-entropy method, we calculate the end-to-end distance 
distribution of the force stretched chain from the moments of the 
distribution, which can be obtained from the extension-force curves 
recorded in single-molecule experiments. If one knows 
force expansion of the extension through the $(n-1)$th power of force, it is 
enough information to calculate the $n$ moments of the distribution. We 
examine the method with three force stretching chain models, 
Gaussian chain, free-joined chain and excluded-volume chain on 
two-dimension lattice. The method reconstructs all distributions 
precisely. We also apply the method to force stretching complex chain 
molecules: the hairpin and secondary structure conformations. We find 
that the distributions of homogeneous chains of two conformations are 
very different: there are two independent peaks in hairpin distribution; 
while only one peak is observed in the distribution of secondary 
structure conformations. Our discussion also shows that the end-to-end distance 
distribution may discover more critical physical information than the 
simpler extension-force curves can give.      
\end{abstract}

\maketitle
\section{Introduction}
\label{sec:intro}
Recent advances in the molecules manipulation have made it possible to 
measure and characterize molecular properties at a single molecule level. 
One of basic characteristics is the extension-force curves 
(EFCs)\cite{smith96,bensimon,zlatanova}. These curves have provided lots 
of interesting and useful physical information about studied 
molecules, going from the detailed elastic properties\cite{smith96} to 
complex structure transitions\cite{rief,bockelmann,liphardt}.  
On theoretical side, many kinds of models have been constructed to 
character and explain the recorded various EFCs of different molecules
\cite{lubensky,cocco,montanari,zhou1,gerland}. 
Except computer simulations, e.g., molecular dynamic or Monte Carlo 
sampling, the calculation of the end-to-end distance distributions 
(EEDDs) of the force stretched molecules is the center problem in using 
statistical mechanical method. In principle, EEDDs can be obtained by 
partition function. But two questions must be faced firstly: one is what 
physical 
interactions should be taken into account; the other is what mathematical 
technique is needed to solve the EEDDs. It is not easy to describe 
physical interactions in complex molecules, such as polyelectrolytes or 
proteins. While specific mathematical technique is not always useful in 
different molecular systems.

In contrast to traditional mind, in this paper we try to extract EEDDs from 
the recorded EFCs in experiments using the maximum-entropy (or least-biased) 
method (MEM). Our motivations are that, first, to our knowledge, little 
concern about EEDDs has been given in previous force stretching models. 
Although EEDDs of simpler molecules may be simpler enough, it is no reason 
to assume that they are still simple for complex molecules, such as secondary 
structure RNA; second, because the EEDDs 
are the results of interplay between intra-molecule and force, they can 
be seen as primitive examining for more realistic physical models. In 
addition, our studies also show that the EEDD at vanishing force calculated 
by MEM keeps almost all characteristic of the exact distribution without force. 
Therefore this method may provide a possible way to directly ``measure" EEDDs 
by force spectroscopy. 

The organization of this paper is as follows. We first, in 
Sec.~\ref{MEMintrod}, 
briefly review the maximum entropy method. In Sec.~\ref{momentsEFC}, the basic 
relations between the distance moments and EFCs are demonstrated.  
In Sec.~\ref{testMEM}, MEM is examined by reconstructing EEDDs of 
three force stretching chain models: Gaussian chain, free-joined chain and 
self-avoiding chain on 2-dimension lattice. We also show in 
Sec.~\ref{applicationRNA} that the method is capable of resolving EEDDs of 
complex chain molecules stretched by force. As an illustration, the model of 
force unzipping double-stranded chain molecules is used to provide  
exact EFCs and EEDDs\cite{liuf}; these results are necessary to calculate and 
compare EEDDs solved by MEM. Section.~\ref{conclusion} is our conclusion. 

\section{Maximum entropy method}
\label{MEMintrod}
Given a finite set of the moments of a distribution function, how to construct 
the function is an old mathematical problem. The MEM has been proved to be useful 
in this problem\cite{med,poland}. From a normalized distribution function 
$P(z)$ on the interval $(0,1)$, the power moments are calculated as  
\begin{eqnarray}
\label{m1}
\mu_n=\int z^nP(z)dz.
\end{eqnarray}
On the other hand, given a set of $(M+1)$ moments, from $\mu_0$ to $\mu_M$, it 
is not always possible to find a positive, well-behave function $P(z)$ that 
will have these moments. The necessary and sufficient conditions for the 
existence of a function $P(z)$ with a set of $(M+1)$ moments on interval 
$(0,1)$ are the Hausdorff relations\cite{med}: 
\begin{eqnarray}
\label{m2}
\sum\limits_{m=0}^k(-1)^m\left(\begin{array}{c}
k\\
m
\end{array}
\right)\mu_{n+m}\ge0&,\hspace{0.4cm}{\rm for}
\ (n,k)=(0,0)\ {\rm to} \ {n+k}\leq M.
\end{eqnarray}

The MEM offers a definite procedure for the construction of the approximate 
distribution $P_M(z)$ based on $(M+1)$ moments as the following 
form\cite{med}:
\begin{eqnarray}
\label{m3}
P_M(z)=\exp\left[-\sum\limits_{n=0}^M\lambda_nz^n\right].
\end{eqnarray}
The $\lambda_n$ are a set of $(M+1)$ constants determined by the
$(M+1)$ known $\mu_n$. This involves a straightforward nonlinear iterative 
procedure that usually converges rapidly\cite{med}.

In general, a real distribution is not always defined on interval $(0,1)$. 
Hence the first step in using MEM is to convert the 
distribution to a function on this interval\cite{poland}. Given that the 
power moments of the original distribution $f(x)$ are $\gamma_m$, and the lower 
and upper bounds are designated as $\alpha_1$ and $\alpha_2$ respectively. 
Defining the extent of the distribution 
\begin{eqnarray}
L=\alpha_2-\alpha_1.
\end{eqnarray}
First, shift the moments $\gamma_m$ to interval $(0,L)$ by  
\begin{eqnarray}
\overline\mu_n=\sum\limits_{m=0}^n(-\alpha_1)^{n-m}\left(\begin{array}{c}
n\\
m
\end{array}
\right)\gamma_m.
\end{eqnarray}
Then scale these moments $\overline\mu_n$ to interval $(0,1)$ by 
\begin{eqnarray}
\mu_n=\overline\mu_n/L^n.
\end{eqnarray}
Thus MEM can be used to calculate distribution $P(z)$. 

Conversely, if the approximate distribution $P(z)$ is solved from moments 
$\gamma_m$, the distribution can be first rescaled from interval 
$(0,1)$ to $(0,L)$ by the change of variable y  
\begin{eqnarray}
g(y)=\frac{1}{L}P\left(\frac{y}{L}\right),\hspace{0.45cm}y\in (0,L).
\end{eqnarray}
Then shift the distribution $g(y)$ to interval $(\alpha_1,\alpha_2)$ 
by 
\begin{eqnarray}
f(x)=g(x-\alpha_1),\hspace{0.45cm}w\in (\alpha_1,\alpha_2).
\end{eqnarray}

\section{Moments from extension-force curves}
\label{momentsEFC}
Assuming that one end of a chain consisting of $N$-links is fixed at 
origin, and external force $f{\bf z_0}$ is exerted on the other end, 
where unit vector ${\bf z_0}$ is along $z$-axis. Let $P_N({\bf R},f)$ 
be the probability distribution function that the end-to-end vector of 
the force stretched chain is ${\bf R}=(R_x,R_y,R_z)$. Then the power 
moments of component $R_z$ distribution $P_N(R_z,f)$ are calculated by   
\begin{eqnarray}
\label{momentexp}
\overline{R_z^m}(f)&=&<({\bf R}\cdot {\bf z}_0)^m>\nonumber\\
&=&\int dR_z (R_z)^m P_N(R_z,f)\\
&=&\int d^3{\bf R} ({\bf R}\cdot {\bf z}_0)^m P_N({\bf R},f).\nomb
\end{eqnarray}
In order to illustrate expressions more seriously and explicitly, 
distribution function of the ideal chains is used\cite{klei,doi} 
\begin{eqnarray}
\label{eeddfunction}
P_N({\bf R},f)&=&{\cal Q}^{-1}[f]\int{\cal D}[{\bf r}(s)]\delta^{3}
\left({\bf R}- \int_{0}^{L}ds{\bf v}(s)\right)\nomb\\
&&\times\exp\left[-\frac{1}{k_BT}\int_{0}^{L}ds
\rho_e({\bf r},s) + \frac{f}{k_BT}{\bf z}_0\cdot \int_{0}^{L}ds{\bf v}(s)
\right],
\end{eqnarray}
where $k_B$ is Boltzmann constant, $T$ is temperature, $L$ is arclength of 
the chain, ``vector" ${\bf r}(s)$ describes the 
local state at arclength point $s$, e.g., in the case of a flexible
Gaussian chain, ${\bf r}$ is a three-dimensional position vector; while 
in case of a wormlike chain, ${\bf r}$ is the unit tangent vector\cite{mak}. 
Vector ${\bf v}(s)$ is also different according to concrete chain model, 
e.g., ${\bf v}=d{\bf r}/ds$ for Gaussian chain, and ${\bf v}={\bf r}(s)$ or 
the tangent vector for wormlike chain. The normalization factor 
${\cal Q}[f]$ is 
\begin{eqnarray}
\label{normform}
{\cal Q}[f]&=&\int d{\bf R}\int{\cal D}[{\bf r}(s)]\delta\left({\bf R}-
\int_{0}^{L}ds{\bf v}(s)\right)\nomb\\
&&\times\exp\left[-\frac{1}{k_BT}\int_{0}^{L} ds\rho_e({\bf r},s) + 
\frac{f}{k_BT}{\bf z}_0\cdot \int_{0}^{L} ds{\bf v}(s)\right]\nomb\\
&=&\int{\cal D}[{\bf r}(s)]\exp\left[-\frac{1}{k_BT}\int_{0}^{L}ds
\rho_e({\bf r},s) + \frac{f}{k_BT}{\bf z}_0\cdot \int_{0}^{L}ds{\bf v}(s)
\right].
\end{eqnarray}
Replacing Eq.~\ref{eeddfunction} into Eq.~\ref{momentexp} and performing 
${\bf R}$ integral we have 
\begin{eqnarray}
\label{eedintegrate}
\overline{R_z^m}(f)&=&{\cal Q}^{-1}[f]\int{\cal D}[{\bf r}(s)]
\left({\bf z_0}\cdot \int_{0}^{L}ds{\bf v}(s)\right)^m\nomb\nomb\\
&&\times\exp\left[-\frac{1}{k_BT}\int_{0}^{L}ds
\rho_e({\bf r},s) + \frac{f}{k_BT}{\bf z}_0\cdot \int_{0}^{L}ds{\bf v}(s)
\right].
\end{eqnarray}
It is easy to prove that Eq.~\ref{eedintegrate} can be rewritten as
\begin{eqnarray}
\label{RzderiveQ}
\overline{R_z^m}(f)=\frac{(k_BT)^m}{{\cal Q}[f]}\frac{\partial^m}{\partial{f^m}}{{\cal Q}[f]}.
\end{eqnarray}
The first moment is just the average extension $Z(f)$ recorded in 
experiments as a given force $f$, 
\begin{eqnarray}
\label{Rzderive1st}
\overline{R_z^1}=Z(f)=\frac{k_BT}{{\cal Q}[f]}\frac{\partial}{\partial{f}}{\cal Q}[f].
\end{eqnarray}
We can alternatively relate the partition function ${\cal Q}[f]$ to 
derivatives of $Z(f)$ with respect to $f$ as follows: 
\begin{eqnarray}
\label{RzQfderive}
\frac{1}{k_BT}\frac{\partial}{{\partial{f}}}Z(f)&=&-\left(\frac{{\cal Q}^{(1)}}
{\cal Q}\right)^2+ \frac{{\cal Q}^{(2)}}{\cal Q},\nomb\\
\frac{1}{k_BT}\frac{\partial^2}{\partial{f^2}}Z(f)&=&2\left(\frac{
{\cal Q}^{(1)}}{\cal Q}\right)^3-3\left(\frac{{\cal Q}^{(1)}}{\cal Q}\right)
\left(\frac{{\cal Q}^{(2)}}{\cal Q}\right)+
\left(\frac{{\cal Q}^{(3)}}{\cal Q}\right),\\
\frac{1}{k_BT}\frac{\partial^3}{\partial{f^3}}Z(f)&=&-6\left(\frac{
{\cal Q}^{(1)}}{\cal Q}\right)^4+12\left(\frac{{\cal Q}^{(1)}}{\cal Q}
\right)^2\left(\frac{{\cal Q}^{(2)}}{\cal Q}\right)\nomb\\
&&-3\left(\frac{{\cal Q}^{(2)}}{\cal Q}\right)^2-4\left(\frac{{\cal Q}^{(1)}}
{\cal Q}\right)\left(\frac{{\cal Q}^{(3)}}{\cal Q}\right)+
\left(\frac{{\cal Q}^{(4)}}{\cal Q}\right),\nomb
\end{eqnarray}
and so on, where
\begin{eqnarray}
\label{deriveQf}
{\cal Q}^{(n)}=\frac{\partial^n}{\partial f^n}{\cal Q}[f].
\end{eqnarray}
According to Eq.~\ref{RzderiveQ}, the moments $\overline{R_z^m}(f)$ can be 
solved in terms of derivatives of $Z(f)$ with respect to force $f$ as follows: 
\begin{eqnarray}
\label{Rzderive}
\overline{R_z^1}&=&Z(f),\nomb\\
\overline{R_z^2}&=&k_BT\frac{\partial}{\partial f}Z(f)+
\left(\overline{R_z^1}\right)^2,\nomb\\
\overline{R_z^3}&=&(k_BT)^2\frac{\partial^2}{\partial f^2}Z(f)-
2\left(\overline{R_z^1}\right)^3+3\left(\overline{R_z^1}\right)
\left(\overline{R_z^2}\right),\\
\overline{R_z^4}&=&(k_BT)^3\frac{\partial^3}{\partial f^3}Z(f)
+6\left(\overline{R_z^1}\right)^4-12\left(\overline{R_z^1}\right)^2
\left(\overline{R_z^2}\right)+3\left(\overline{R_z^2}\right)^2
+4\left(\overline{R_z^1}\right)\left(\overline{R_z^3}\right).\nomb
\end{eqnarray}
The above relations show that if one has the first $(n-1)$ derivatives of 
$Z(f)$, then this is enough information to calculate the first $n$ moments 
of the distribution $P_N(R_z,f)$. These derivatives of $Z(f)$ can be 
obtained by expanding the extension $Z(f)$ in a Taylor series 
about the reference force $f_0$ as 
\begin{eqnarray}
\label{taylor}
Z(f)=Z(f_0)+\frac{\partial}{\partial f}{Z(f_0)}{\Delta f}
+\frac{1}{2}\frac{\partial^2}{\partial f^2}{Z(f_0)}{\Delta f}^2+
\frac{1}{6}\frac{\partial^3}{\partial f^3}{Z(f_0)}{\Delta f}^3,
\end{eqnarray}
where 
\begin{eqnarray}
\Delta f=f-f_0.
\end{eqnarray}
In general, no analytical $Z(f)$ is used in real situation; only the 
EFCs are recorded in experiments. All the derivatives 
have to be calculated by numerical methods. 

Before beginning the next section, we clarify our procedure: 
first calculate different derivatives of $Z(f)$ from EFCs by numerical method; 
then use Eq.~\ref{Rzderive} to obtain necessary power moments of 
$P_N(R_z,f)$; and finally, apply MEM presented in Sec.~\ref{MEMintrod} to 
construct approximate EEDDs.

\section{test of MEM: three force stretching chain models}
\label{testMEM}

In this section, the MEM is examined with three force stretching chain models 
which have different statistical properties: Gaussian chain, free-joined chain 
and excluded-volume (EV) chain on two dimension. The main reason to choice 
these  
model is that their EEDDs with forces have exact expressions. EEDDs 
and EFCs of the models are solved by statistical mechanical methods firstly. 
Then seeing the obtained EFCs as experiment data, approximate distributions is 
computed by MEM according to procedure mentioned in above section. Distributions 
solved by two methods are compared finally. 

For each chain model, EEDDs at three nonzero forces are calculated 
respectively. In addition, distributions at zero force are also 
solved. Because that EEDD without force is important in polymer 
research, such as the calculation of root-mean-square end-to-end distance, 
it is interesting to see whether the distributions calculated by MEM at zero 
force can keep main characteristic of the exact EEDD. Though our method only 
solve $P_N(R_z,0)$, the length distribution $P_N({\bf R},0)$ could be obtained 
from the numerical relations provided by Domb et al. early\cite{domb}. 

\subsection{Gaussian model}
\begin{figure}[htpb]
\begin{center}
\includegraphics[width=0.4\columnwidth]{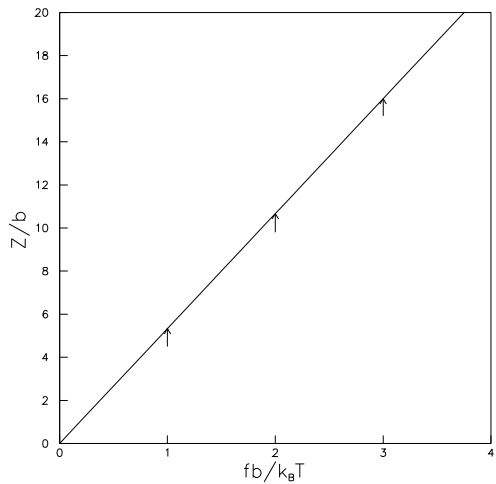}
\caption{EFC of the Gaussian chain, here $N=16$. The three 
arrows point out forces in which corresponding EEDDs are calculated by MEM.  } 
\label{gaussplotefc}
\end{center}
\end{figure}

As the simplest $N$-link chain model, the energy with force $f{\bf z_0}$ in 
Gaussian model\cite{doi} is expressed as  
\begin{eqnarray}
\label{gausseng}
{\cal E}=\frac{3k_BT}{2b}\int_{0}^{L=Nb} ds\left( \frac{\partial {\bf r}}
{\partial s}\right)^2 - f{\bf z}_0\cdot\int_{0}^{Nb}ds\frac{\partial {\bf r}}
{\partial s}, 
\end{eqnarray}
where $b$ is effective bond length, ${\bf r}$ is position vector. Using path 
integral method\cite{klei,doi,zhou}, the EEDD of Gaussian chain 
stretched by force is derived as 
\begin{eqnarray}
\label{gaussReed}
P_N({\bf R},f)=\left(\frac{3}{2\pi Nb^2}\right)^{3/2}
\exp\left\{-\frac{3}{2Nb^2}\left({\bf R}-
\frac{Nb^2f{\bf z_0} }{3k_BT}\right)^2\right\}.
\end{eqnarray}
Correspondingly, the component $R_z$ distribution in ${\bf z_0}$ direction 
can be integrated by 
\begin{eqnarray}
\label{gaussRzeed}
P_N(R_z,f)&=&\int dR_xdR_yP_N({\bf R},f)\nomb\\
&=&\left(\frac{1}{2\pi Nb^2}\right)^{3/2}
\exp\left\{-\frac{3}{2Nb^2}\left(R_z-\frac{Nb^2f}{3k_BT}
\right)^2\right\}.
\end{eqnarray}
The extension versus force then is calculated as
\begin{eqnarray}
\label{gaussefc}
Z(f)=\int dR_zR_zP_N(R_z,f)=\frac{Nb^2}{3k_BT}f.
\end{eqnarray}
As an illustration, we choice $N=16$ and plot the function in 
Fig.~\ref{gaussplotefc}. This function will be viewed as EFCs ``measured" 
in experiments.  

\begin{figure}[htpb]
\begin{tabular}{cc}
\includegraphics[width=0.4\columnwidth]{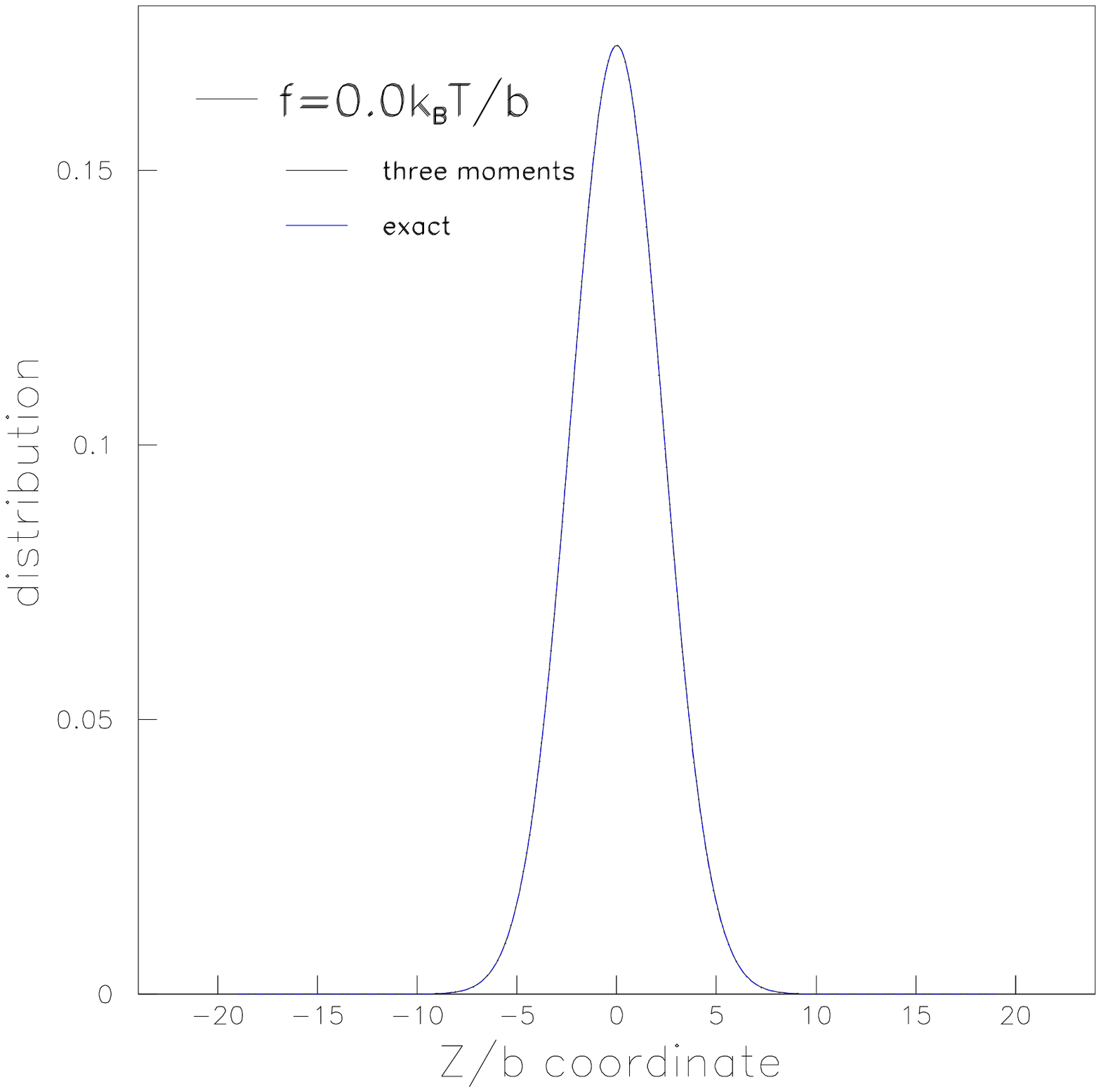}&
\includegraphics[width=0.4\columnwidth]{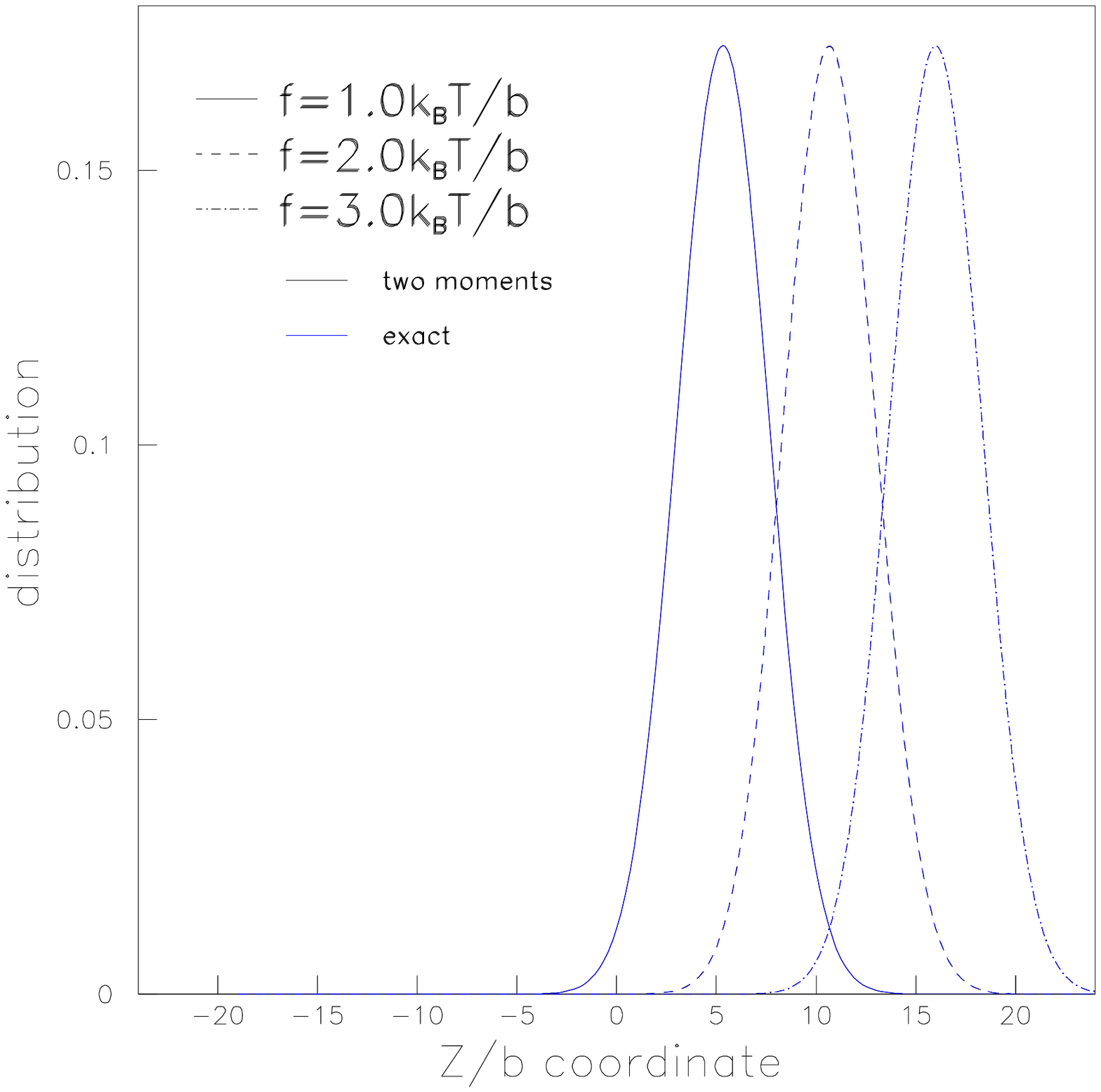}\\
(a)&(b)
\end{tabular}
\caption{ Comparing EEDDs solved by MEM (the black lines) with exact EEDDs 
calculated by Eq.~{\protect\ref{gaussRzeed}} (the blue lines) for force 
stretching Gaussian chain model: (a) $f=0.0 k_BT/b$; (b) $f=1.0, 2.0$ and 
$3.0 k_BT/b$. Here three moments approximation is used in MEM. Overlapping 
of two color lines demonstrates that the MEM can precisely reconstructs the 
exact EEDDs of Gaussian chain at any force value. }
\label{gaussploteed}
\end{figure}

Before applying MEM, firstly expand Eq.~\ref{gaussefc} in Taylor series about 
force $f_0$ as  
\begin{eqnarray}
\label{gaussefcexp}
Z(f)=Z(f_0)+\frac{Nb^2}{3k_BT}(f-f_0),
\end{eqnarray}
or $Z(f_0)=Nb^2f_0/3k_BT$ and $\partial{Z}/\partial{f_0}=Nb^2/3k_BT$. 
Three moments can be obtained at any given force $f_0$ through 
Eq.~\ref{Rzderive} directly. Then approximate distributions are solved by 
MEM. Three distributions calculated by MEM and their comparing with exact 
Eq.~\ref{gaussRzeed} at forces $0.0,1.0,2.0,3.0k_BT/b$ are shown in 
Fig~.\ref{gaussploteed}. Considering that the extension is linear with force, 
three moments approximation is used in this model. Obviously, MEM can precisely 
reconstruct distributions of the Gaussian chain. In fact, because the approximation 
function $P_2(R_z,f)$ is just the Gaussian distribution, it is not unexpected that 
MEM reconstruct EEDDs of Gaussian chain perfectly. In addition, EEDD at 
$f=0.0k_BT/b$ is the same with distributions at nonzero forces, since the 
first-order derivate of $Z(f)$ is constant at any force, 

\subsection{Free-joined chain model}
Free-joined chain model has been used to fit the observed EFCs of  
force stretching single-stranded DNA experiments\cite{smith96}. The 
model is defined as a chain with N-link of length $b$ in which all 
rotational angles occur with equal probability\cite{klei,doi}. When 
exerted external force $f{\bf z_0}$ on one end of the chain, the force 
potential energy is written as  
\begin{eqnarray}
\label{fjcforceng}
{\cal E}_f=-f{\bf z_0}\cdot\sum\limits_{n=1}^{N} {\bf r_n}, 
\end{eqnarray}
where $r_n$ are bond vectors with constant length $|r_n|=b$. According to  
Eq.~\ref{eeddfunction}, the distribution function $P_N({\bf R},f)$ of the 
end-to-end vector ${\bf R}$ is 
\begin{eqnarray}
\label{fjcReed}
P_N({\bf R},f)&=&\frac{\exp\left[\beta f{\bf z_0}\cdot{\bf R}\right]P_N({\bf R},0)}
{\int d{\bf R}\exp\left[\beta f{\bf z_0}\cdot{\bf R}\right]P_N({\bf R},0)
},\nomb\\ 
\end{eqnarray}
where $\beta=1/k_BT$, and $P_N({\bf R},0)$ is the EEDD without applied force, 
which has been given in literature\cite{klei} 
\begin{eqnarray}
\label{fjcnoforce}
P_N({\bf R},0)&=&\frac{1}{ 2^{N+1}(N-2)!\pi b^2R}\sum\limits_{n=0}^{[(N-R/b)/2]}
(-1)^n\left(\begin{array}{c}
n\\
N
\end{array}
\right)
(N-2n-R/b)^{N-2},
\end{eqnarray}
where $R$ is the length of vector $\bf R$. The normalization factor or the partition function ${\cal Q}[f,T]$ can 
be calculated exactly as 
\begin{eqnarray}
\label{fjcpart}
{\cal Q}[f,T]&=&\int d{\bf R}P_N({\bf R},f)\nomb\\
&=&\left(\frac{4\pi}{\beta fb}\sinh\left({\beta fb}\right)\right)^N.
\end{eqnarray}

\begin{figure}[t]
\begin{center}
\includegraphics[width=0.4\columnwidth]{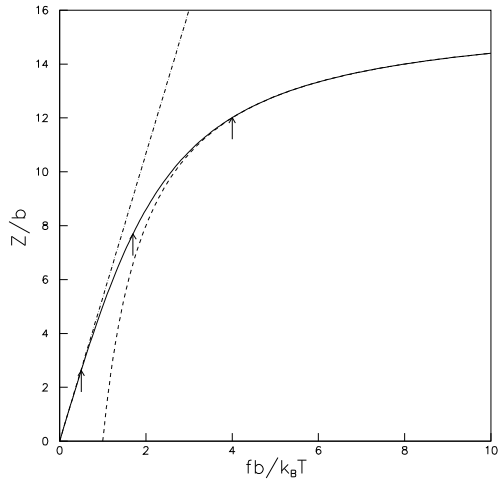}
\caption{EFC of free-joined chain, here $N$=16. The 
dot-dash and dash curves are asymptotic curves corresponding to large and 
small forces respectively. Three arrows point out different nonzero forces 
in which corresponding EEDDs are calculated by MEM.  }
\label{fjcplotefc}
\end{center}
\end{figure}

\begin{figure}[htpb]
\begin{tabular}{cc}
\includegraphics[width=0.4\columnwidth]{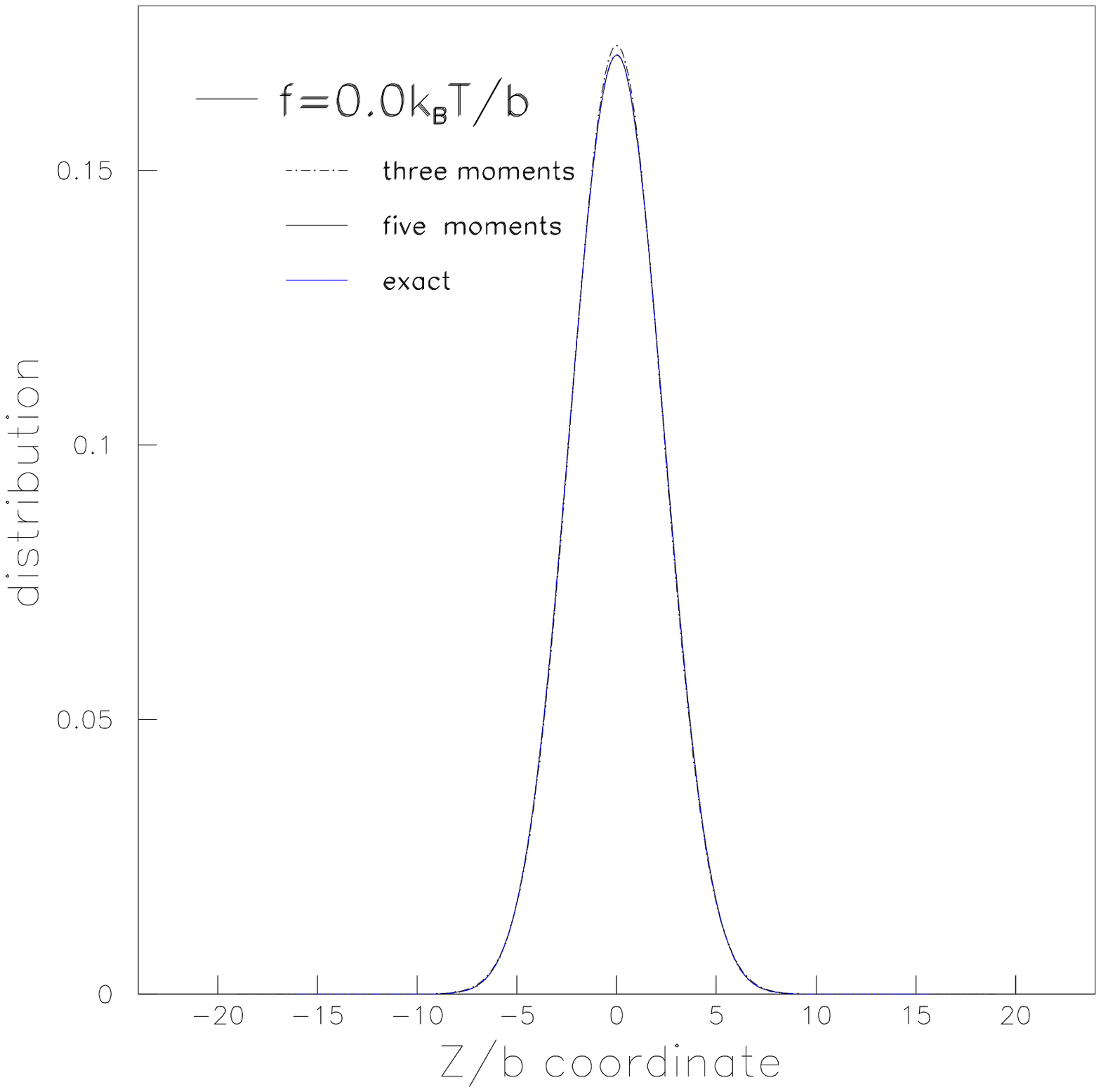}&
\includegraphics[width=0.4\columnwidth]{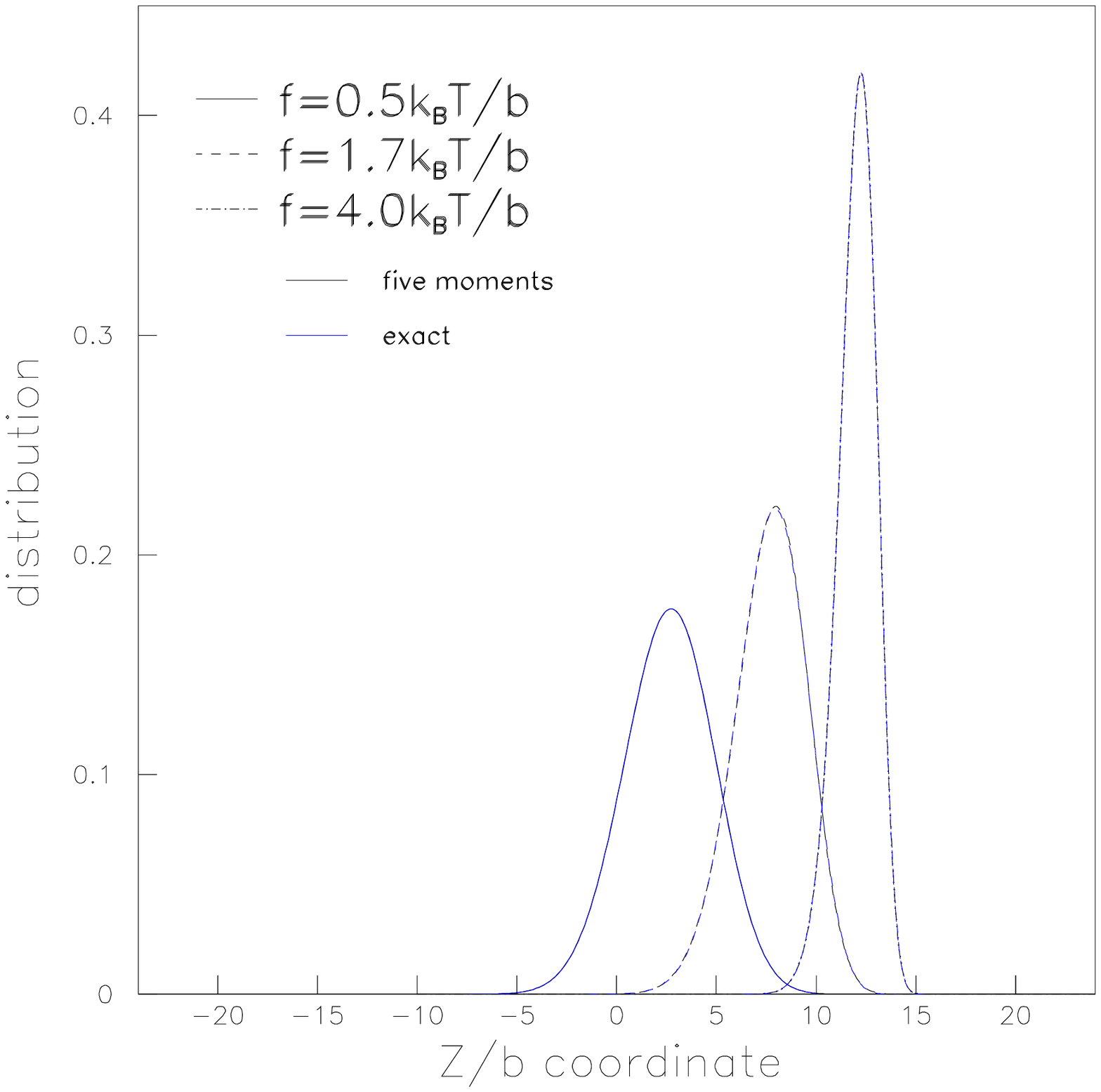}\\
(a)&(b)
\end{tabular}
\caption{ Comparing EEDDs solved by MEM (the black lines) with exact EEDDs 
given by Eq. ~{\protect\ref{fjcRzeed}} (the blue lines) for force 
stretching free-joined chain model: (a) $f=0.0k_BT/b$. We calculate the 
EEDDs using three (black dash line) and five moments (black solid line), 
respectively. The EEDD of five moments slightly derives from the 
distribution of three moment at origin, which is confirmed by exact EEDD. 
(b) $f=0.5, 1.7$ and $4.0k_BT/b$. Here five moments are necessary. 
Unlike in Gaussian chain, not only the maximum values of EEDDs are movable, 
but also the distribution regions are variable at different forces. 
Overlapping of two color lines show that MEM can very precisely 
reconstruct the EEDDs of free-joined chain at any given force. 
}
\label{fjcploteed}
\end{figure}

Then EEDD of component $R_z$ can be obtained by integral of Eq.~\ref{fjcReed} with respect to components 
$R_x$ and $R_y$ as 
\begin{eqnarray}
\label{fjcRzeed}
P_N(R_z,f)=\frac{\exp\left(\beta f R_z\right)}{\left[4\pi \sinh\left(\beta f b 
\right)/{\beta fb}\right]^N} P_N(R_Z,0),
\end{eqnarray}
where
\begin{eqnarray}
\label{fjcRzeednoforce}
P_N(R_z,0)=\frac{1}{2^{N}(N-2)!b}\int_{R_z/b}^{N}dx\sum\limits_{0}^{[(N-x)/2]}
(-1)^n\left(\begin{array}{c}
n\\
N
\end{array}
\right)
(N-2n-x)^{N-2},\hspace{0.4cm}{\rm R_z\ge0}.
\end{eqnarray}
The average extension $Z(f)$ is then given as    
\begin{eqnarray}
\label{fjcefc}
Z(f)&=&bN\left[-\frac{1}{\beta fb}+\coth\left(\beta fb\right)\right].
\end{eqnarray}
Eq.~\ref{fjcefc} is served as experiment data to check our MEM. As an example, 
take N=$16$, and its EFC is shown in Fig.~\ref{fjcplotefc}.
We expand Eq.~\ref{fjcefc} in Taylor series at different forces,   
$0.0,0.5,1.7$, and $4.0k_BT/b$. EFC at these forces has different 
asymptotic formula; see dash curves in Fig.~\ref{fjcplotefc}. 
Similarly to the case of of Gaussian model, EEDDs calculated by MEM and 
their comparing with exact distribution are shown in Fig.~\ref{fjcploteed}. 
At $f=0.0k_BT/b$, because the third-order derivative of $Z(f)$ is not zero, EEDDs of 
three and five moments are solved by MEM respectively; see 
Fig.~\ref{fjcploteed}(a). The distributions of three and 
five moments are slightly different at origin: EEDD of three moments is the same 
with the distribution of Gaussian model; while distribution value at origin 
calculated by five moments is smaller, which is the same with prediction of exact 
EEDD. Our results show that MEM is sensitive enough to detect the fine 
difference of EEDDs from simple EFCs. When force is nonzero, EEDDs solved by 
MEM are the same with exact EEDDs.  

\subsection{Self-avoiding chain model}
As a more realistic model, the self-avoiding chain which accounts for 
EV interactions plays very important role in polymer 
theory\cite{doi,flore}. But in force stretching problem, almost all 
theoretical models implicated that EV interaction can be negligible. This 
assumption is doubted at small force region. In this section, We try to 
simulate the force stretched EV chain of $N$-link as $N$-step self-avoiding 
walks (SAW) on two dimensional (2D) quadratic lattice. The early work of 
Domb {\it{et al.}} has demonstrated that EEDD of self-avoiding chain differ 
appreciably from Gaussian distribution\cite{domb}. It is interesting to see 
weather MEM can recover the EV properties exactly when the force tends to zero. 

First we formulate the force stretching partition function ${\cal Q}[f,T]$ and 
EEDD $P_N(n,f)$ as follows 
\begin{eqnarray}
\label{sa-2}
{\cal Q}[f,T]=\sum_{n=-N}^{n=+N}C_N^x\left(n\right)\exp\left(f\beta n\right),
\end{eqnarray}
and  
\begin{eqnarray}
\label{sa-3}
P_N(n,f)=\frac{C_N^x\left(n\right)\exp\left(f\beta n\right)}
            {{\cal Q}\left[f,T\right]}, 
\end{eqnarray}
where $C_N^x\left(n\right)$ is the number of walks whose final $x$ coordinates 
are $n$. Extension function $Z(f)$ then can be calculated from EEDD accurately. 

\begin{figure}[htpb]
\begin{center}
\includegraphics[width=0.4\columnwidth]{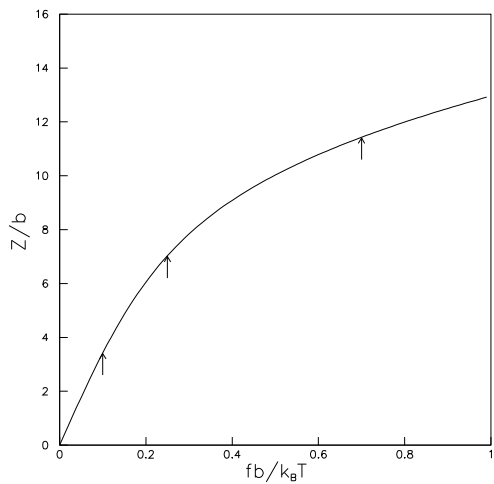}
\caption{EFC of SAW chain, here $N=20$. Three arrows point 
out different forces in which corresponding distributions are calculated 
by MEM.} 
\label{sawfz}
\end{center}
\end{figure}

\begin{figure}[htpb]
\begin{tabular}{cc}
\includegraphics[width=0.4\columnwidth]{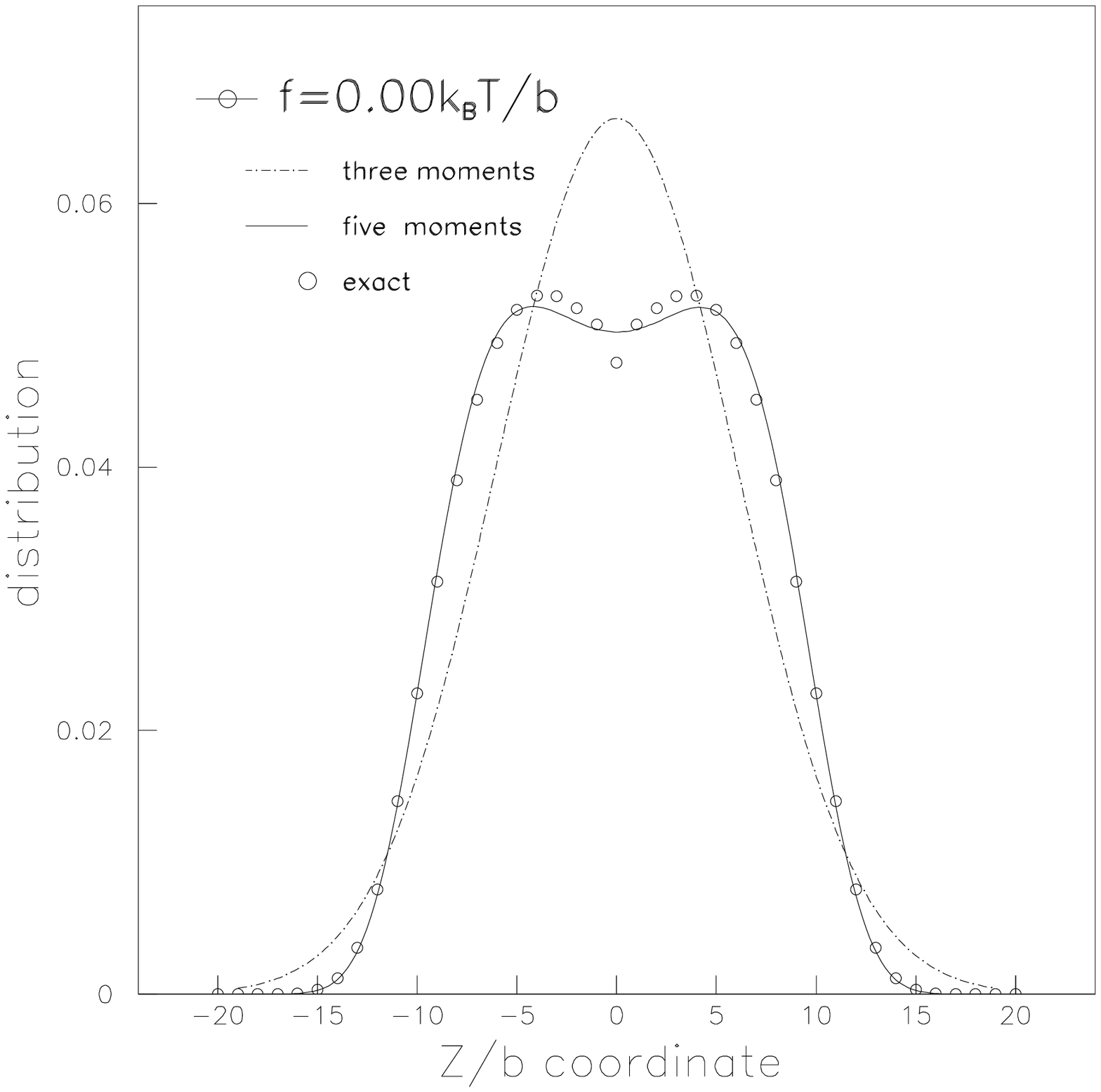}&
\includegraphics[width=0.4\columnwidth]{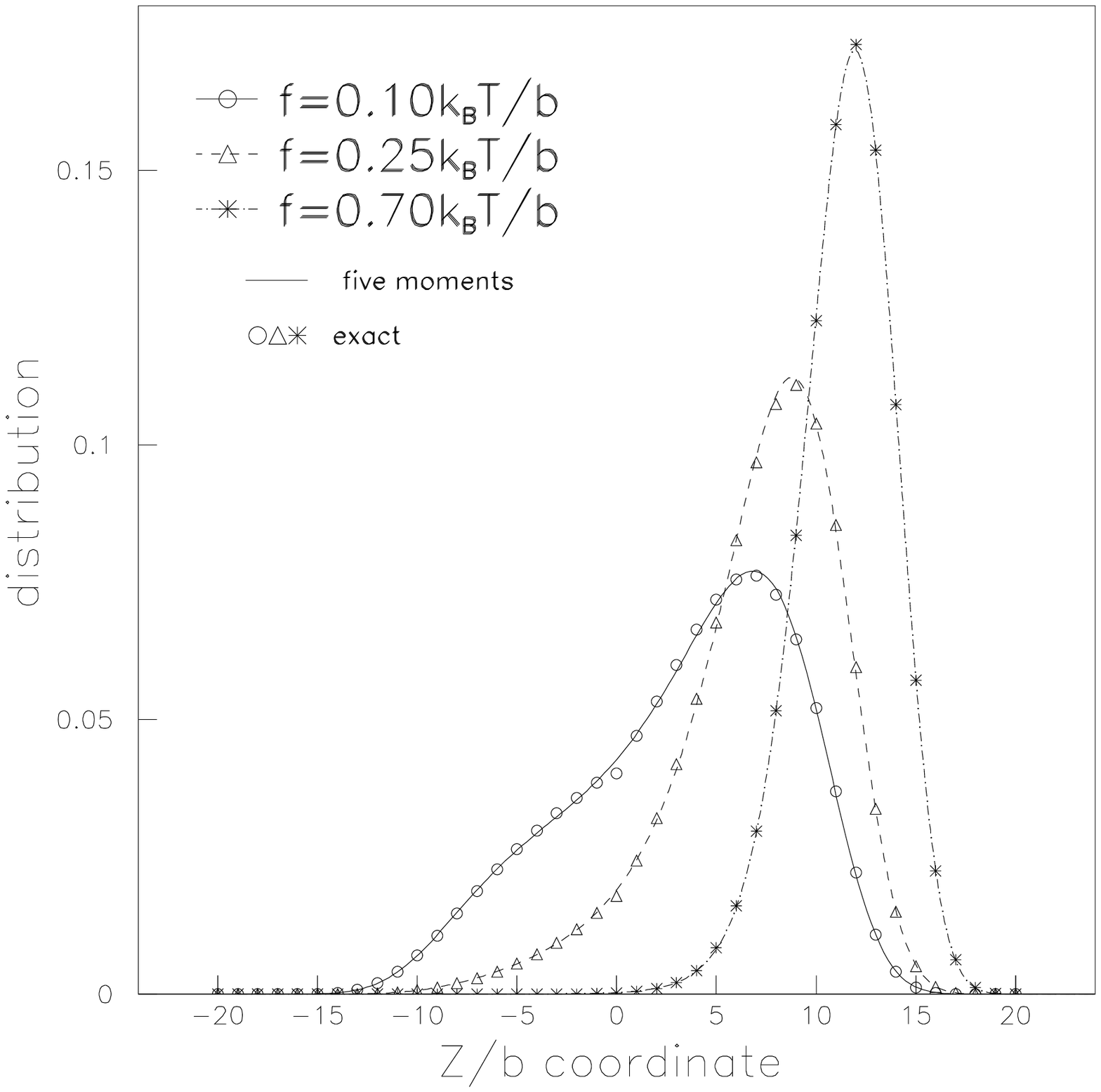}\\
(a)&(b)
\end{tabular}
\caption{ Comparing EEDDs solved by MEM (lines) with exact EEDDs (symbols) 
given by Eq.~{\protect\ref{sa-3}} for force stretching self-avoiding chain. 
(a) $f=0.00k_BT/b$. We calculate EEDD using three and five moments 
respectively. Unlike EEDDs of Gaussian chain or free-joined chain at $f=0.$, 
two peaks in EEDD obtained by five moments (solid lines) appear in this model, 
which are demonstrated by exact enumeration (circles). (b) $f=0.10, 0.25$ and 
$0.70k_BT/b$. Five moments are necessary. Good fitness of the lines and symbols 
shows that EEDDs calculated by MEM can recover real distribution precisely.}
\label{saweed}
\end{figure}

As an illustration, we exactly enumerate all $20$-step SAWs on 2D lattice. 
According to Eq.~\ref{sa-2}, we calculate EFC and plot it in 
Fig.~\ref{sawfz}.   
The numerical expansions of extension of the chain at 
forces $0.0,0.10,0.25$, and $0.70k_BT/b$ are calculated respectively. Then 
using MEM, EEDDs at these forces are solved; see Fig.~\ref{saweed}. At 
force $0.0k_BT/b$, distributions of three and five moments are different 
apparently. Domb {\it{et al.}} have pointed out that instead of Gaussian 
distribution, the distributions considering EV effect on 2D lattice can be 
well fitted by a function form of $\exp(-\abs x^4)$, which will be seen that 
the portion of the EEDD 
near to the origin is more ``flat-topped", and the decay of distribution for 

larger values of x is sharper\cite{domb}. EEDD calculated by five moments at 
zero force precisely recovers these major aspects. However, it is unexpected 
that even a slight dip in the value of distribution can be recovered by the 
MEM; see Fig.~\ref{saweed}(a). Because the origin of the dip arises from the 
restriction of no returns to the origin\cite{domb}, the result demonstrates 
again that the MEM is very sensitive to detect the fine structure restrictions 
from the simple EFC. This characteristic in distribution is still preserved 
at small forces, such as at force $0.10k_BT/b$ in Fig.~\ref{saweed}(b). 
   
From the analysis of self-avoid chain, we conclude that EV may play important 
role even in force stretching problem. Especially, EV effect can be reflected 
more explicitly from EEDD, instead of simple EFCs. 

\section{EEDDs of force stretching complex molecules: hairpin and 
secondary structure conformations}
\label{applicationRNA}
From the deduction of Eq.~\ref{RzQfderive}, the relations are independent of 
interactions between the units in a chain. On the 
other hand, units of any real polymer always interact with each other, 
e.g., the simple electrostatic repulsion of the phosphodiester backbone of 
DNA, and complex hydrophobic interaction in proteins. Hence it is valuable 
to see what MEM can tell us about the interactions in molecule. 
Because recent mechanical single molecular 
experiments have turn their attentions to molecular structure transitions 
induced by force, such as dsDNA or ssDNA (RNA) force 
unzipping\cite{bockelmann,liphardt}, it is natural to apply MEM to these 
experiment data firstly. In particular, the EEDDs at critical force is of 
interest\cite{liphardt}. However, considering that the MEM is 
very sensitive to the shapes of EFCs, and current experiment data 
are not fine enough, in this paper we do not ready to apply our method in 
experiment data directly. In this section we will make use of EFCs 
solved by an theoretical model of force stretching hairpin and secondary 
structures conformations in 2D plane\cite{liuf} as ``experiment" data. 
Because our model also provide the exact EEDDs, comparing with EEDDs 
derived by MEM will ensure the availability of MEM when our method is 
applied to real scenario in future. In following section, we first give 
a brief overview about the statistical model of force stretching 
chain molecules of hairpin and secondary structure conformations. 
The details of the model are given 
elsewhere\cite{liuf}. 

\subsection{A statistical mechanical model of force stretching 
chains of hairpin and secondary structure conformations}
Hairpin and secondary structure conformations are the basic 
models for antiparallel $\beta$-sheet in protein and RNA molecules\cite{chen1,chen2}. The partition function ${\cal Q}_N(T;f)$ of a 
$(N+1)$ monomer ((N+1)-mer) chain molecules stretched by force $f$ is 
formulated as 
\begin{eqnarray}
\label{partitionSS}
{\cal Q}_N(T;f)=\sum_E\sum_\Delta g_N(E;\Delta)e^{-\beta(E-f\Delta)},
\end{eqnarray}
where $\Delta$ is end-to-end distance (EED) of the chain along force 
direction, and $g_N(E;\Delta)$ is the number of conformations having 
energy $E$ and EED $\Delta$. Because the energy contributed by force is 
only related with EEDs, we divide any conformation of the chain into two 
parts: one is main chain (MC), in which does not involve any contacts; 
the other is nested regions (NRs), which form hairpins, loops and 
turns. If only one NR is allowed in conformations, they are named hairpin, 
otherwise secondary structure conformations. On 2D 
lattice, the nested regions contribute $\pm1$ or $0$ to whole EED value 
according their outmost contacts directions\cite{liuf}. The $g_N(E;\Delta)$ 
is simplified as a multiplication of the number of 
conformations of MC and NRs, i.e., Eq.~\ref{partitionSS} can be rewritten 
as  
\begin{eqnarray}
\label{partitionNOW}
{\cal Q}_N(T;f)=\sum_n\sum_E\sum_\Delta C^{MC}(n,\Delta)C^{ NRs}(n,E)
e^{-\beta (E-f\Delta)},
\end{eqnarray}
where $n$ is the number of unrelated NRs in conformations, 
$C^{MC}(n,\Delta)$ and $C^{NRs}(n,E)$ are the number of conformations of MC 
and NRs respectively. For hairpin conformations $n=1$.   

Because our model is restricted on 2D lattice, the values of 
$C^{MC}(n,\Delta)$ at given $n$ can be counted exactly by enumeration and  
extrapolation method\cite{liuf}. Whereas calculation of $C^{NRs}(n,E)$ is modified and 
extended from nested polymer graph theory (NPGT) developed by Chen and Dill 
recently\cite{chen1,chen2}. The idea behind the NPGT is that the number of 
conformations of any arrangement of NRs is a product of each number of 
conformations of each NR restricted by EV requirement. In NPGT, 
different arrangement of NRs is represented by polymer graph, the 
diagrammatic representations of intrachain contacts, and each unrelated NR 
can be independently seen as a polymer graph or subpolymer graph. So the 
calculation of the number of conformations for any given subpolymer graph is 
the centra of NPGT. According to the NPGT, the number of conformations of any 
subpolymer graph having $m$ subunits is a product of matrices: 
\begin{eqnarray}
\label{graphsum}
{\bf U}\cdot{\bf S}_{t_m}\cdot{\bf Y_{t_mt_{m-1}}}\cdot{\bf S}_{t_{m-1}}\cdots{{\bf S}_{t_1}}\cdot{\bf U}^{t},
\end{eqnarray}
where ${\bf U}=\{1,1,1,1\}$, ${\bf U}^{t}$ is the transpose of
${\bf U}$, ${\bf S}_i$ is structure matrix of $i$th subunit, and $Y_{ij}$ is 
viability matrix\cite{chen1,chen2}.  

\begin{figure}[htpb]
\begin{center}
\includegraphics[width=0.4\columnwidth]{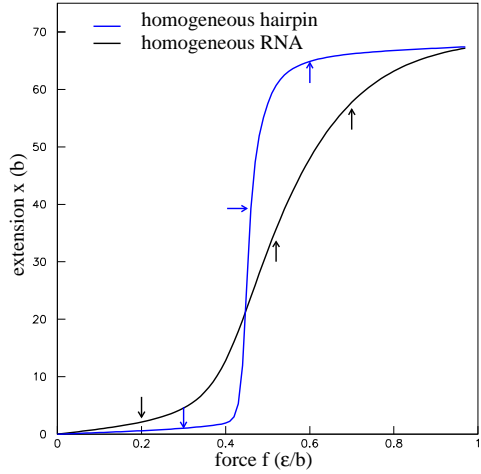}
\caption{EFCs of 70-mer homogeneous chains of hairpin (blue 
lines) and secondary structure conformations (black line). Here temperature is 
$0.28\varepsilon/k_B$. Six arrows point out different forces in which 
corresponding distributions of two conformations are calculated by MEM.} 
\label{fzss}
\end{center}
\end{figure}

We obtain EEDD from the partition function 
${\cal Q}(T;f)$, 
\begin{eqnarray}
\label{eedsshp}
P_N(\Delta,f)=e^{\beta f\Delta}\times
\sum_n\sum_E C^{MC}(n,\Delta)C^{NRs}(n,E) e^{-\beta E}/
{{\cal Q}_N(T;f)},
\end{eqnarray}
and the average extension function $Z(f)$ is calculated exactly from above 
EEDDs. For comparing, we calculate EFCs of 70-mer homogeneous 
chains of hairpin and secondary structure conformations; see Fig.~\ref{fzss}. 
Here the homogeneous chain means that any contact of 
two monomers in chain contributes energy $-\varepsilon$ ($\varepsilon>0$). 
Considering to the importance of 
sequence in secondary structure molecules, we also give EFC of a 70-mer 
specific sequence, A$\cdots$ACCCCCU$\cdots$UC$\cdots$CAAAAAG$\cdots$G,  
where the dots represent 15-mer A, U, C and G respectively; see 
Fig.~\ref{desefc}(a). In contrast to homogeneous chain, only A-U or 
C-G pair contributes energy $-\varepsilon$. 

\subsection{Single- and multipeak distributions}
\begin{figure}[htpb]
\begin{tabular}{cc}
\includegraphics[width=0.4\columnwidth]{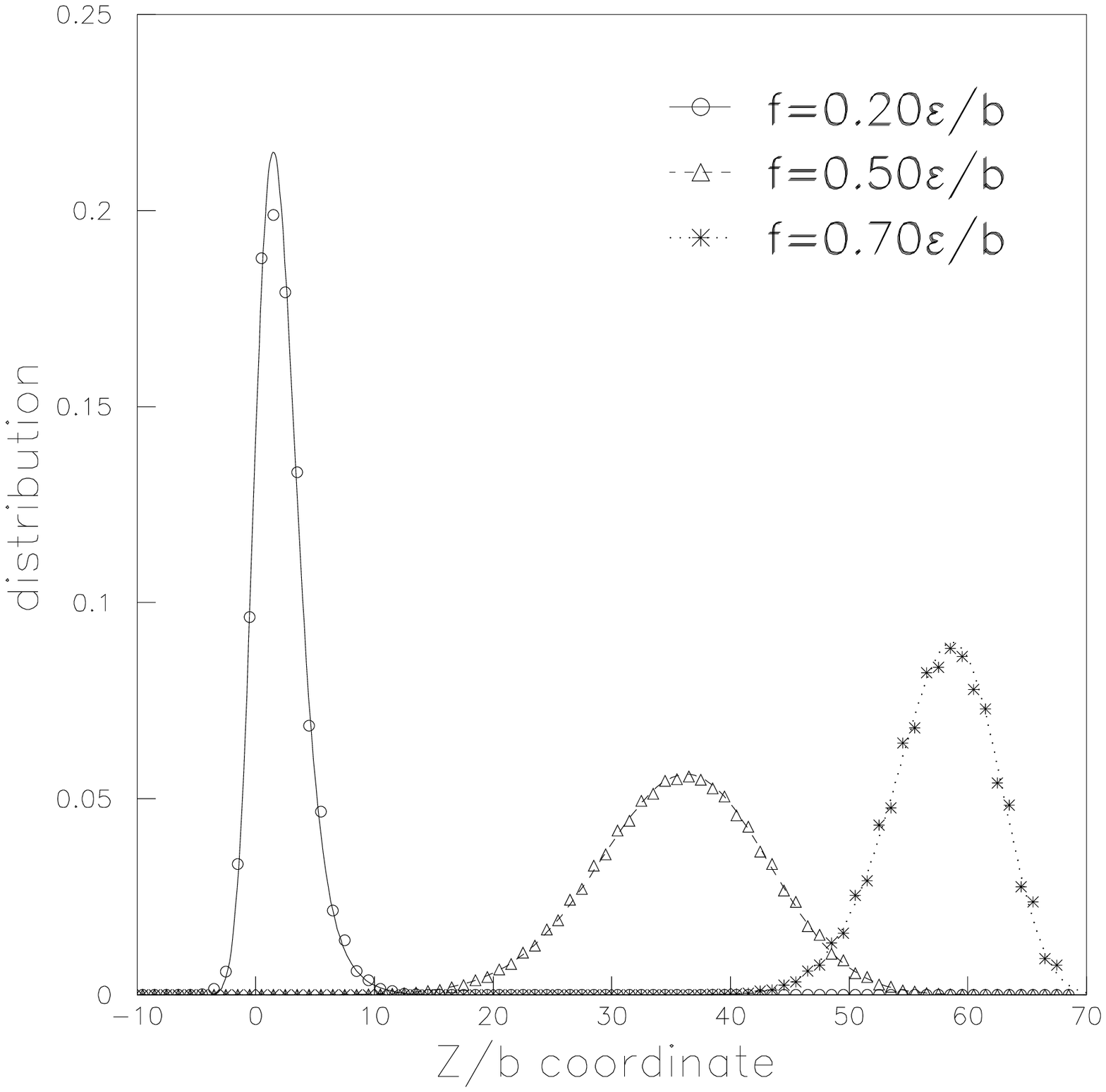}&
\includegraphics[width=0.4\columnwidth]{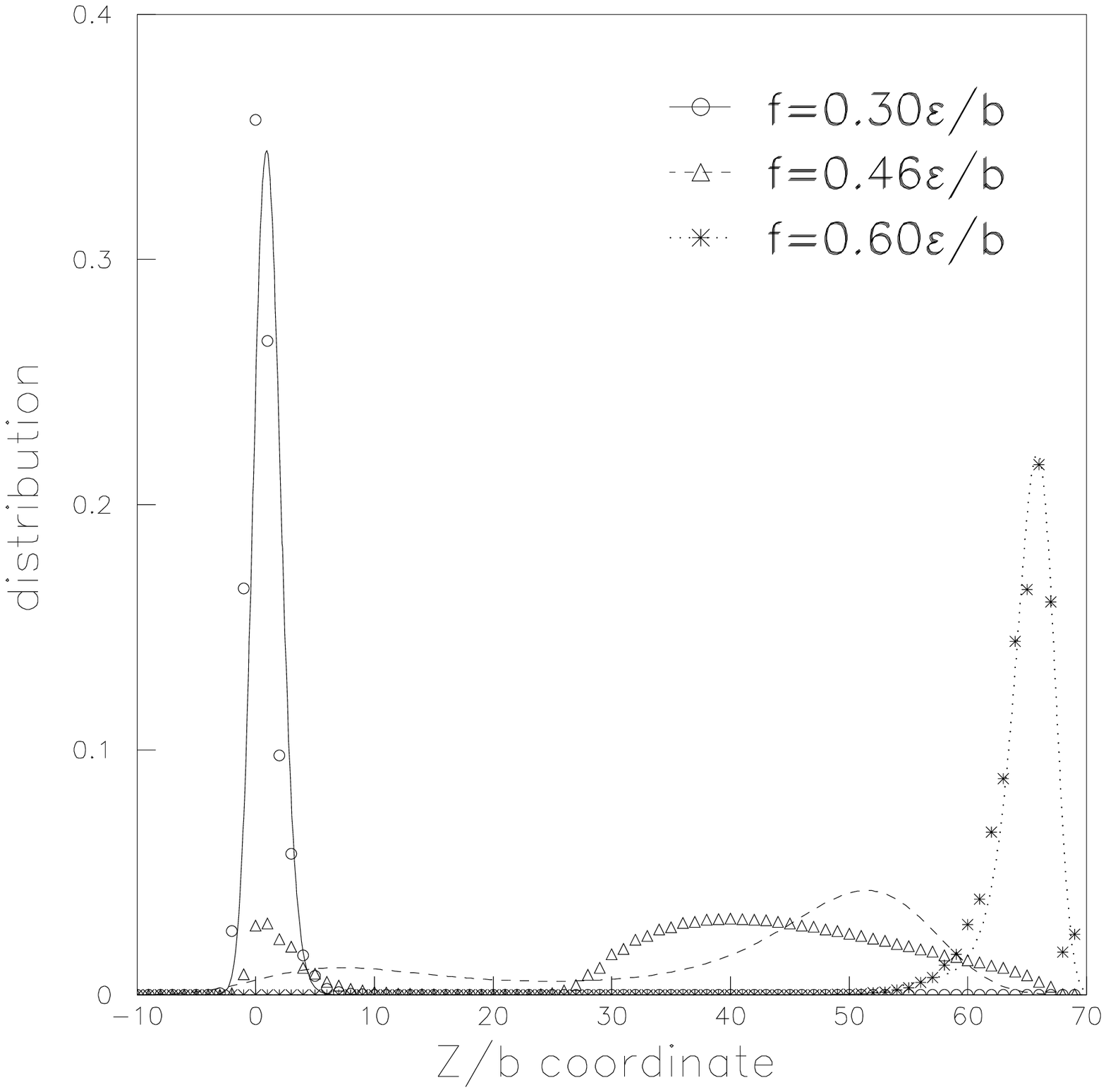}\\
(a)&(b)
\end{tabular}
\caption{ Comparing between EEDDs solved by MEM (lines) and exact EEDDs 
given by Eq.~{\protect\ref{eedsshp}} (symbols) for homogeneous chains. 
(a) Secondary structure conformations, where $f=0.20$, $0.50$ and 
$0.70\varepsilon/b$. (b) Hairpin conformations, where $f=0.30$, $0.46$ and 
$0.60\varepsilon/b$. Five moments are necessary. Two independent peaks in 
EEDD at force $0.46\varepsilon/b$ appear in hairpin model (triangle and 
dash line), while it does not present in secondary structure conformations. } 
\label{cphpAss}
\end{figure}

\begin{figure}[htpb]
\begin{tabular}{cc}
\includegraphics[width=0.4\columnwidth]{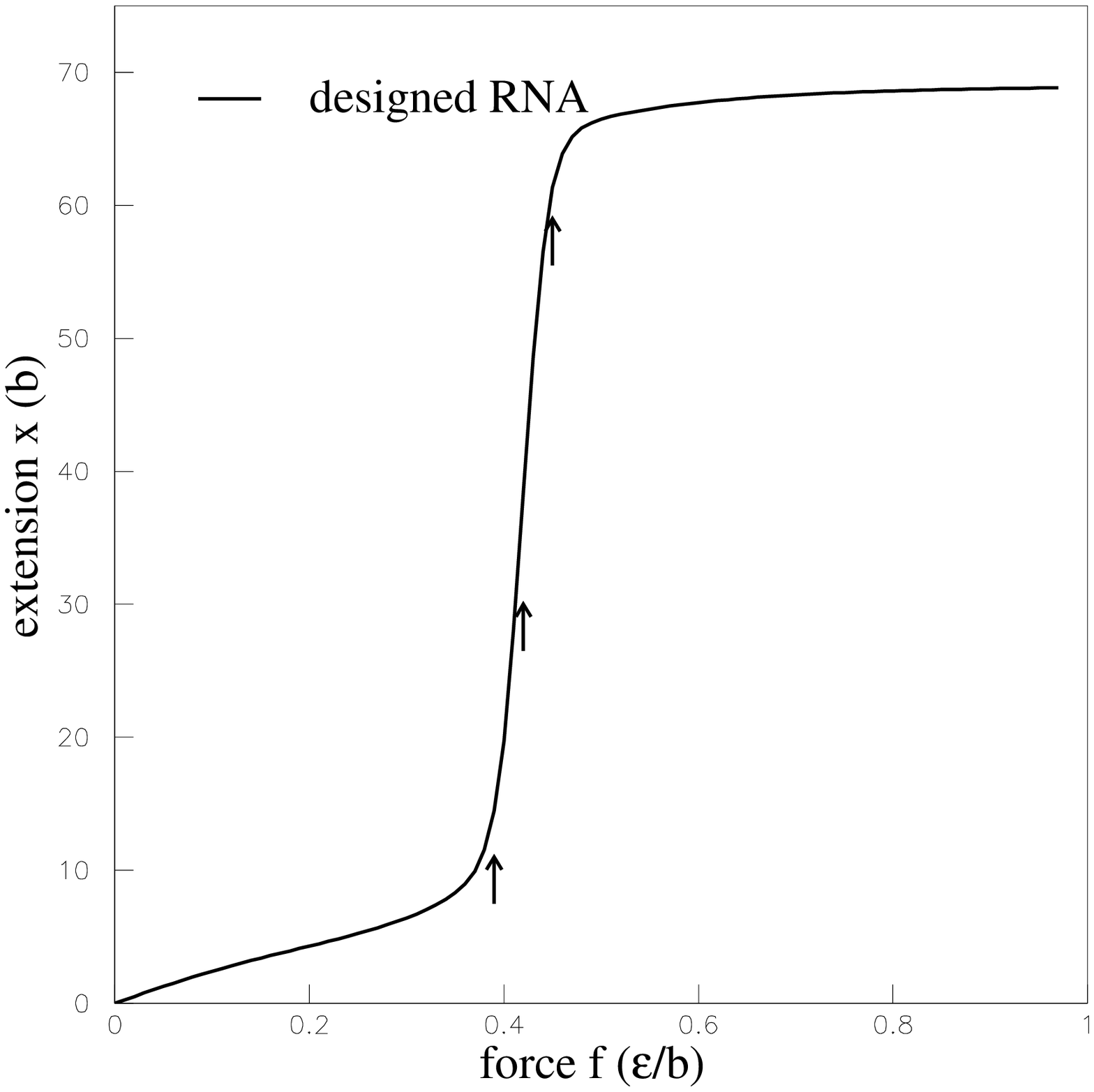}&
\includegraphics[width=0.4\columnwidth]{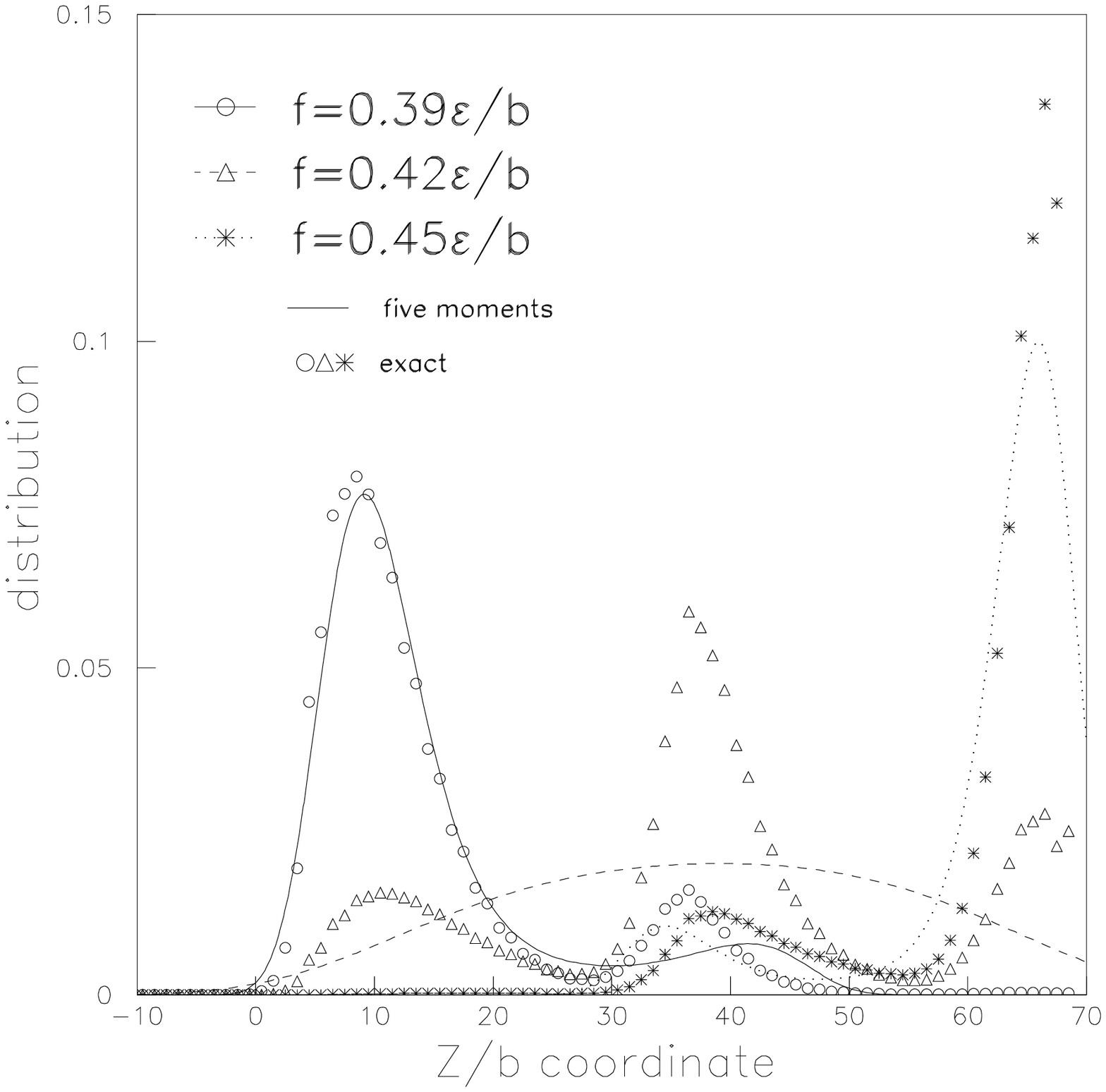}\\
(a)&(b)
\end{tabular}
\caption{(a) EFC of 70-mer specific sequence of   
secondary structure conformations, where temperature is $0.28\varepsilon/k_B$. 
Three arrows point out different forces in which corresponding 
distributions are calculated by MEM. (b) Comparing EEDDs solved by MEM 
(lines) with exact EEDDs (symbols), where $f=0.39$, $0.42$ and $0.45\varepsilon/b$. Five moments are necessary. Unlike EEDDs of homogeneous chains, 
three peaks appear in the exact EEDD at $f=0.42\varepsilon/b$ (triangles), 
while the EEDD calculated by MEM only shows an abnormal expansion at this force 
(dash line). }
\label{desefc}
\end{figure}

To be the same with previously section, we compute all EEDDs by numerical 
expansion of EFCs at different forces for different chain molecules; see 
Figs.~\ref{cphpAss} and \ref{desefc}(b). 

The shapes of the EEDDs of the complex molecules are very different from 
those of simple molecules observed in Sec.~\ref{testMEM}. The most obvious 
feature is that the distribution regions of the complex chains expand in middle 
force whereas shrinking at smaller and larger forces.  It is results of 
attracted interaction between monomers. Secondly, it 
seems that EFCs of homogeneous chains of hairpin and secondary structure 
conformations are similar except that extensions increase slowly or fast, 
however, the EEDDs calculated by MEM are completely different: only one peak 
is observed in the distribution at any force in secondary structure 
conformations; see Fig.~\ref{cphpAss} (a); while in hairpin conformations, 
two peaks located at shorter and longer EEDs present during narrow force range 
(between 0.42 to 0.49$\varepsilon/b$), but no conformations 
with other EEDs in between; see Fig.~\ref{cphpAss} (b). In addition, EEDDs 
of specific sequence show more complex shapes. At $f=0.39\varepsilon/b$ 
the distribution has two peaks, then they quickly fuse into one peak with 
a small force increasing 0.03$\varepsilon/b$, and finally the peak separates 
into two peaks at $f=0.45\varepsilon/b$. To explore the phenomena of 
single- or multipeak, comparing with the exact EEDD is essential. These results 
are plotted in corresponding figures. We find that the two peaks in 
distributions predicted by MEM are consistent with exact EEDDs of specific 
sequence and homogeneous hairpin chain. At force 
$0.42\varepsilon/b$, however the exact distribution of specific sequence 
appears three peaks, whereas MEM predicts one peak only.  

According to Eq.~\ref{RzQfderive}, the first-order derivative of the 
average extension $\overline{R_z^1}(f)$ with respect of force $f$ can also 
be written as $d\overline{R_z^1}/df=(\overline{R_z^2}-\overline{R_z^1}^2)
/k_BT$.  
The formula is the same with the definition of heat capacity $C(T)$ except 
that the energy and temperature are replaced by EED and force, respectively. We 
believe that the EED plays important roles in force stretching chains 
problem, which is very similar with the roles played by energy in thermal 
melting biomolecules, at least in nucleic acids\cite{liuf}. Many useful 
insights can be given through this analogy. Since the energy distribution 
can reveal molecular structure transitions induced by heating\cite{poland}, 
the EEDD might discover structure transitions driven by force. E.g., 
the EEDDs in Fig.~\ref{cphpAss} show that the transitions in secondary 
structure and hairpin conformations are ``one-state" and ``two-state", 
respectively. These terms are borrowed from thermal melting case\cite{poland,
tostesen}. The transition difference exhibited in two conformations warns us 
that the simpler EFCs may cover critical physical information; the 
investigation of EEDDs is necessary to determine physical properties in 
force ``melting" chain molecules. 

In traditional theory, physical properties of polymers only relate with the 
number of monomers, such as cooperativity or melting transition type\cite{doi,flore}. 
But in biomolecules, the monomer sequence may affect physical results dramatically. 
The apparent case in EEDDs is the number of peaks of 
the specific sequence in Fig.~\ref{desefc}, though its EFC is simpler. The 
case of specific sequence warns us again that EFCs may be too simple to 
obtain real and useful information about the studied molecules. 
Because five moments cannot reconstruct three peaks in 
distribution\cite{poland}, we only solved a quick expansion in the 
distribution region. To explore three or more 
peaks in EEDDs, higher moments are necessary. Although our MEM fails to 
predict three peaks, the abnormal expansion of the distribution 
observed by MEM between two forces arising two peaks still can be seen 
as a sign of appearing of multipeak.  

\section{Conclusions}
\label{conclusion}
In this paper, contrary to the traditional mind, we calculate end-to-end 
distance distribution of force stretching chain molecules from the measured 
EFCs by using MEM. Because the method is independent of polymer energy formula  
 or structure details, it provide a useful and simple way 
to detect the real physical information about complex molecules. Many 
results, such as the important role played by EV interactions, single- or 
multipeak in EEDDs can be obtained from the simple EFCs. 
It is interesting to see whether these results can be found in real 
stretching experiments. 

\begin{acknowledgments}
It is a pleasure to acknowledge Dr. H.-J. Zhou,  Y. Zhang and 
Prof. H.-W. Peng for many helpful discussions in this work.
\end{acknowledgments}


\begin{thebibliography}{99}

\bibitem{smith96} 
S. B. Smith, Y. J. Cui, and C. Bustamante, \sci{271}{795}{1996}.

\bibitem{bensimon}
B. Maier, D. Bensimon, and V. Croquette, \pnas{97}{12002}{2000}.

\bibitem{zlatanova}
J. Zlatanova, S. M. Lindsay, and S. H. Leuba, Prog. Biophs. Mol. Biol. 
{\bf 74}, 37 (2000).

\bibitem{rief} 
M. Rief, H. Clausen-Schaumann and H. E. Gaub, \nats{6}{6}{1999}.

\bibitem{bockelmann}
U. Bockelmann, B. Essevaz-Roulet, V. Viasnoff, and F. Heslot, 
\bio{82}{1537}{2002}.

\bibitem{liphardt}
J. Liphardt, B. Onoa, S.B., Smith, I.J. Tinoco, and C. Bustamante, 
\sci{292}{733}{2001}.

\bibitem{lubensky}
D. K. Lubensky and D. R. Nelson, \pre{65}{031917}{2002}, and references
therein.

\bibitem{cocco}
S. Cocco, R. Monasson, and J. F. Marko, \pnas{98}{8608}{2001}.

\bibitem{montanari}
A. Montanari and M. M$\acute{e}$zard, \prl{86}{2178}{2001}.

\bibitem{zhou1}
H.J. Zhou, Y. Zhang, and Z.-C. Ou-Yang, \prl{114}{8694}{2001}.

\bibitem{gerland}
U. Gerland, R. Bundshuh and T. Hwa, \bio{81}{1324}{2001}.

\bibitem{liuf}
F. Liu, L.R. Dai and Z.-C. Ou-Yang, (2002), cond-mat/0212268.

\bibitem{med} 
L. R. Mead and N. Papanicolaou, \jcp{\bf 25}, {2404} (1984).

\bibitem{poland} 
D. Poland, \jcp{\bf 112}, {6554} (2000).

\bibitem{klei} 
H. Kleinert, {\it Path Integrals in Quantum Mechanics, Satistics,and Polymer 
Physics} (World Scientific, Singapore, 1990).

\bibitem{doi} 
M. Doi and S. F. Edwards, {\it The Theory of Polymer Dynamics}, 
(Clarendon Press, Oxford, 1986).

\bibitem{mak} 
J. F. Marko and E. D. Siggia, \mol{28}{8759}{1995}.

\bibitem{zhou} H. J. Zhou, Z. Yang and Z. C. Ou-Yang, \pre{\bf 62}, {1045} 
(2000). 

\bibitem{flore} 
P. J. Flory, {\it Principles of Polymer Chemistry} (Cornell Univ. Press, 
Ithaca, 1953).

\bibitem{domb}
C. Domb, J. Gills and G. Wilmers, Proc. Phys. Soc. (London) {\bf85}, 
625 (1965). 

\bibitem{chen1}
S.-J. Chen and K.A. Dill, \jcp {\bf 103}, {5802} (1995).

\bibitem{chen2}
S.-J. Chen and K.A. Dill, \jcp {\bf 109}, {4602} (1999). 

\bibitem{tostesen}
E. T$\o$stesen, S.-J. Chen, and K. A. Dill, \jpc{105}{1618}{2001}.

\end{thebibliography}
\end{document}